# Solving Combinatorial Optimization Problems on Fujitsu Digital Annealer

Yu–Ting Kao
*Graduate Institute of Applied Physics,
National Chengchi University
B.S. Program in Electrophysics,
National Chengchi University
Department of Money and Banking,
National Chengchi University*
Taipei City, Taiwan
107102007@nccu.edu.tw

Jia-Le Liao
*B.S. Program in Electrophysics,
National Chengchi University
Department of Computer Science,
National Chengchi University*
Taipei City, Taiwan
110703048@nccu.edu.tw

Hsiu–Chuan Hsu
*Graduate Institute of Applied Physics,
National Chengchi University
Department of Computer Science,
National Chengchi University*
Taipei City, Taiwan
hcjhsu@nccu.edu.tw

**Abstract**

**Combinatorial optimization problems are ubiquitous in various disciplines and applications. Many heuristic algorithms have been devoted to solve these types of problems. In order to increase the efficiency for finding the optimal solutions, an application–specific hardware, called digital annealer (DA) has been developed for solving combinatorial optimization problems using quadratic unconstrained binary optimization (QUBO) formulations. In this study, we formulated the number partitioning problem and the graph partitioning problem into QUBO forms and solved such problems with the DA developed by Fujitsu Ltd. The QUBO formulation of the number partitioning problem was fully connected. The DA found the overall runtime for the optimal solution to be less than 30 seconds for 6500 binary variables. For the graph partitioning problem, we adopted modularity as the metric for determining the quality of the partitions. For Zachary's Karate Club graph, the modularity obtained was 0.445, a 6% increase against D–wave Quantum Annealer and Simulated Annealing. Moreover, to explore the DA's potential applications to real–world problems, we used the search for communities or virtual microgrids in a power distribution network as an example. The problem was formulated into graph partitioning. It is shown that the DA effectively identified community structures in the IEEE 33–bus and IEEE 118–bus network.**

## I. Introduction

Combinatorial optimization problems are ubiquitous in modern society, such as parcel delivery [1], workforce assignment [2] and building optimal power systems [3]. However, many combinatorial optimization problems belong to the complexity class of NP-hard. Several heuristic algorithms have been developed for solving these problems. One well–known and widely adopted method is Simulated Annealing (SA), inspired by the annealing process in metallurgy. It facilitates the approximation of global optima within extensive search spaces of optimization problems. While it often yields approximate solutions to the global minimum, these approximations prove to be sufficient for numerous practical scenarios [4].

In recent years, quantum annealing [5][6] has caught significant attention from interdisciplinary communities as quantum technology improves rapidly. Quantum annealing is essentially based on the well–known physics model, the Ising model, that has been used to demonstrate important physical concepts. It has been shown that many combinatorial optimization problems can be converted to solving for the ground state configuration of the Ising models. A practical problem that has a direct analogy to the Ising model is number partitioning [7][8]. The number partitioning problem can be converted to a fully-connected Ising model and is hard to solve when the ratio between the number of bits b used for storing each element and the number of elements N satisfy $\frac{b}{N} \geq 1 - \log(\frac{2^N}{2N})$. This issue has attracted physicists' attention and the complexity has shown to possess a phase transition that separates the easy and hard-to-solve regimes [7][8]. The quantum annealing approach has been implemented with superconducting quantum bits to solve such problems. D–wave is the first commercialized hardware performing quantum annealing and has been adopted for solving complex Ising models [9][10] and real-world problems [6]. However, the current quantum processors still experience noise, intermediate scale and limited connectivity. The advantage of quantum annealing over simulated annealing is thus still under investigation with careful scrutiny [5][11].

Beside simulated and quantum annealing, the Digital Annealer (DA) [12][13] has been developed by Fujitsu Ltd. It is an application–specific CMOS hardware designed to perform SA. Despite the similarity to SA, Fujitsu DA possesses three major distinct features to speed up the optimization process. First, DA initiates all runs from the same arbitrary state, rather than starting each run from a random state, to avoid calculations of the initial energy repeatedly. Second, it employs a parallel–trial scheme where at each Monte Carlo step, the flip of each variable is performed in parallel. Third, DA utilizes an escape mechanism to reduce time spent in local minima. If no flip is accepted in a specific Monte Carlo step, subsequent acceptance probabilities are artificially increased by a dynamically controlled variable. This mechanism enables the algorithm to overcome narrow barriers in the optimization landscape. Thus, as a result of the combination of the specific hardware architecture and the modifications in the algorithm, DA is able to perform efficient exploration in large solution space for complex optimization problems.

Combinatorial problems are pervasive in the real world. Among them, one type of problem, called community detection or graph partitioning, has its application in virtual microgrid detection of power distribution networks. Owing to the emergence of renewable energy, the conventional centralized power distribution network requires an upgrade to adapt to various electricity generation and transmission methods. Virtual microgrids provide a new solution to the upgrade plan by reducing power loss during transmission, attaining higher efficiency and having better compatibility with green energy [14][15].

In our work, we investigate number partitioning and graph partitioning problems using Fujitsu Digital Annealer (DA). We also utilized D-Wave Simulated Annealing Sampler to conduct SA on our problems as a benchmark. Moreover, we study community detection and apply it to the partition of virtual microgrids in power distribution systems. This paper is organized as follows: the QUBO formulation and overview are given in section II, the results obtained with Fujitsu DA are presented in section III, and lastly, the conclusion of this study is given in section IV.

## II. FORMULATION

### A. Quantum Ising model

The foundation of DA is to solve for the configuration of a set of binary variables with the lowest cost function. This problem can be understood as solving for the lowest–energy state of Ising models in quantum physics,

$$E(S) = -\sum_{i,j} w_{i,j}\, s_i s_j - \sum_i h_i s_i \qquad (1)$$

where $s_i \in \{-1,1\}$ is the i-th binary spin variable, $w_{i,j}$ is the interaction between the spins at site $i$ and $j$, $h_i$ is the bias term or the local magnetic field, and S is the array of all the binary variables. A wide range of combinatorial optimization problems can be mapped to (1). Nonetheless, (1) can be converted to another representation with binary variables $\{0,1\}$ by defining the variable $b_i = \frac{s_i + 1}{2}$. The formulation is dubbed as the Quadratic Unconstrained Binary Optimization (QUBO) formulation. In Fujitsu DA, the input objective function uses the QUBO form. Below, we use number partitioning to demonstrate the conversion into the QUBO formulation.

The number partitioning problem is a well–known and extensively studied problem with various formulations. In its most commonly discussed version, the goal is to partition a given set of numbers into two subsets, A and B, such that the difference between the sums of A and B is minimized. Considering a set of numbers, $S = \{S_0, S_1, S_2, \ldots S_{m-1}\}$, we introduce binary variables $x_j$, where $x_j = 1$ if $S_j$ is assigned to set A, and $x_j = 0$ if $S_j$ is assigned to set B. The sum of set A is given by $SUM_A = \sum_{j=0}^{m-1} S_j X_j$ and the sum of set B is $SUM_B = \sum_{j=0}^{m-1} S_j - \sum_{j=0}^{m-1} S_j X_j$. The difference between the two sums, denoted by D, is

$$D = |\,SUM_B - SUM_A\,| = |\,c - 2\sum_{j=0}^{m-1} S_j X_j\,| \qquad (2)$$

where $c = \sum_{j=0}^{m-1} S_j$ is the sum of the numbers in S.

The QUBO form is obtained by squaring (2), resulting in $D^2$ as a quadratic form that approaches zero when D is minimized. Thus, the objective function for minimization is

$$D^2 = c^2 + 4X^T Q X \qquad (3)$$

where $X = (x_0, x_1, \ldots x_{m-1})^T$, $Q_{ii} = S_i(S_i - c)$ and $Q_{ij} = Q_{ji} = S_i S_j$ [16].

In the next two subsections, we present the QUBO formulation for graph partitioning and electrical grid partitioning.

### B. Graph Partitioning

Graph partitioning or community detection is to identify subgraphs in a given graph $G(V, E)$. The metric for evaluating the quality of community detection is modularity, proposed by Newman and Girvan [17]. The modularity Q for weighted graphs is given in [18] as below,

$$Q = \frac{1}{2m} \sum_{i,j} (A_{ij} - \gamma \frac{k_i k_j}{2m}) \delta(c_i, c_j) \qquad (4)$$

where $m = \frac{1}{2} \sum_{i,j} A_{ij}$ is the weighted total number of edges, $A_{ij}$ is the coefficient in row i and column j of the adjacency matrix of a graph, $\gamma$ is the resolution parameter which is taken to be 1 in the study, $k_i$ is the weighted degree of node $i$, and $\delta(c_i, c_j) = 1$ if node i and node j are of the same community otherwise 0. The first term in (4) computes the fraction of weighted edges in the community and the second term computes the fraction of weighted edges if the vertices are connected randomly. The maximum value of the modularity is 1. The larger the modularity, the denser the sub–graphs compared to random connectivity in that sub-graph. Thus, the objective for community detection is to maximize modularity.

To formulate the problem into an Ising model, we let $x_i = (x_{i0}, x_{i1}, x_{i2}, \ldots, x_{ik}, \ldots x_{i(K-1)})$ be the vector associated with node i and let $x_{ik}$ be a binary variable, where K is the total number of communities. If node i is in group k, $x_{ik} = 1$ and all the other binary variables in vector $x_i$ are zero. Thus, the objective function for DA is written as

$$M = -X^T Q X \quad (5)$$

where the vector $X = (x_{00}, x_{01}, x_{02}, \ldots, x_{ik}, \ldots x_{n-1,K-1})^T$. Besides the objective function, we still need two constraints. The first is that each node must be allocated to only one group,

$$\sum_{k=0}^{K-1} x_{ik} = 1, \quad i = 0, 1, \ldots, n-1. \quad (6)$$

The second constraint is that each group has at least one node to avoid empty groups,

$$\sum_{i=0}^{n-1} x_{ik} \geq 1, \quad k = 0, 1, \ldots, K-1. \quad (7)$$

### C. Electrical modularity

For a real–world case study, we apply DA to perform the search of virtual microgrids in power distribution networks, where the weights of edges are determined by the inverse of the absolute value of impedance [19]. The larger the weights, the less the power loss for the edges. We used two well–known test cases, IEEE 33–bus [20] and IEEE 118–bus [21]. The resistance r and reactance x of each edge were given in the database. The weights were calculated by $\frac{1}{|r+jx|}$. Both IEEE 33–bus and IEEE 118–bus were imported from the Python package PandaPower version 2.13.1 [22], and converted to NetworkX graphs using the "create_nxgraph" method. With respect to IEEE 118–bus, we excluded parallel edges in our analysis.

In our study, we harnessed Fujitsu's one–way one–hot constraint method that sped up the solution exploration to impose (6). We utilized Fujitsu DA's inequality constraint separation feature to impose (7) and set the coefficient for each inequality "lambda" to 1 for partitioning Karate Club, IEEE 33–bus, and IEEE 118–bus graphs.

## III. RESULTS AND DISCUSSION

### A. Number partitioning problem

The overall runtime of digital annealing was explored. Additionally, the test cases were solved with simulated annealing [23]. The SA was executed on a local machine with an Intel i7-1165G7 @ 2.80GHz processor. We prepared six different sets of integers to investigate whether array size had a distinct impact on the overall runtime. The experimental data consisted of six sets, each containing a varying number of integers ranging from 4000 to 6500 with an increment of 500. Each integer ranged from 1 to 10000. These were generated randomly using Python's random library.

Across different array sizes, the minimum energy values obtained by DA and SA were [0, 0, 0, 1, 1, 0]. However, it was found that for the smallest array size 4000, SA required more trial initial states in order to achieve the optimal solution. The result indicated that the distinct features of DA enabled it to compute optimal answers straightforwardly. The total runtime on DA was less than 30 seconds for 6500 numbers and scale linearly with array size.

Because the number of bits required to represent the coefficient in the QUBO form was smaller ($\log_2^{10000} \approx 13$) than the array size, this experiment belonged to the easy-to-solve category [8]. The efficiency and solving power of DA will be explored with a higher bit precision and more complex test cases in our future work.

### B. Graph partitioning – Karate club

Zachary's Karate club network describes social interactions among members in a karate club at an American university [24]. The Karate Club graph was imported from the Python library NetworkX version 3.1 [25]. The modularity against number of communities is shown in Fig. 1. The two curves were obtained by DA and SA respectively, with SA executed on a local machine with an Apple M1 processor. DA shows a higher modularity for all numbers of communities. Moreover, it shows that the partition of four communities yields the best modularity, in agreement with previous studies. Interestingly, the largest modularity (0.445) obtained by DA is larger than that obtained from previous studies (0.42) with quantum annealing [26][27]. For four communities, the partitioning result is shown in Fig. 2.

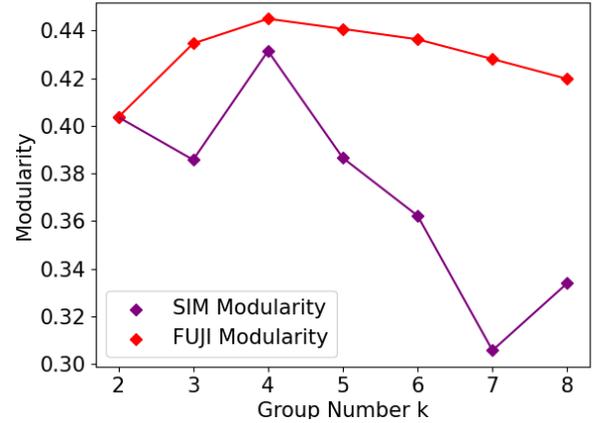

Fig. 1. Karate Club Modularity Versus Group Number.

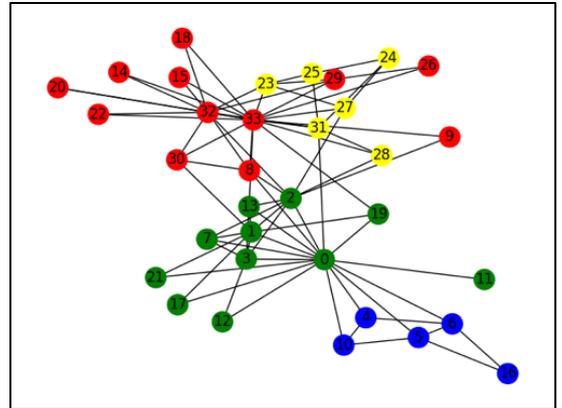

Fig. 2. Partitioned Graph with the highest modularity 0.445, with group number K=4, by Fujitsu Digital Annealer

## C. Application – Electrical virtual micriogrids

We applied our method to IEEE 33, IEEE 118–bus test cases for community detection. The modularity against total number of communities for IEEE 33–bus is shown in Fig. 3. The largest modularity is 0.74 for K=7. The partitioning result for K=7 is shown in Fig. 4. The edges on the boundaries between virtual microgrids have weaker weights. The average weight of the boundaries is 0.88 $\Omega^{-1}$, while the average weight of the rest of the edges is 2.14 $\Omega^{-1}$. The partitioning result shows that the nodes within each community are more densely connected than the nodes across different communities. Moreover, we computed the partition for IEEE 118–bus using DA. The modularity against total number of communities is shown in Fig. 5. Each point was obtained with the parameter "time_limit_sec" set to 10 on DA. The largest number of communities computed is 35, which requires 4130 binary variables. For this problem scale, considerable time was given to SA for manually tuning the penalty coefficients but less than optimal results were obtained, and the runtime on SA grew significantly as the number of communities increased. In stark contrast, DA gave us an optimal solution within 12 seconds by courtesy of its various features. DA indicates that the largest modularity is 0.8196 for K=11, which the partitioning result is shown in Fig. 6.

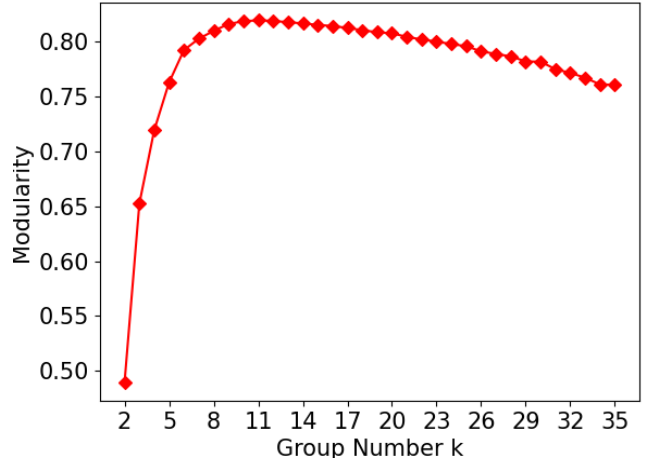

**Fig. 5.** Fujitsu IEEE 118–bus Modularity Versus Group Number

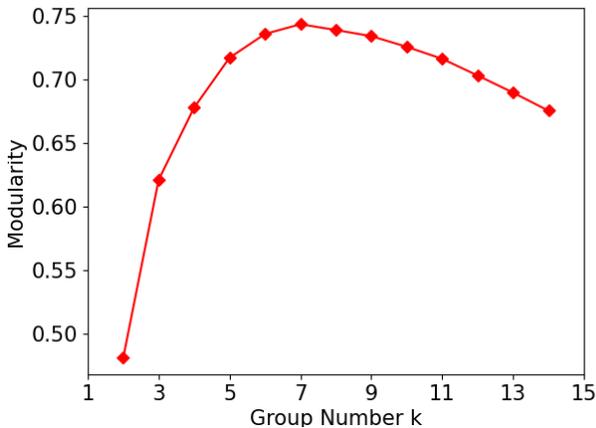

**Fig. 3.** Fujitsu IEEE 33–bus Modularity Versus Group Number

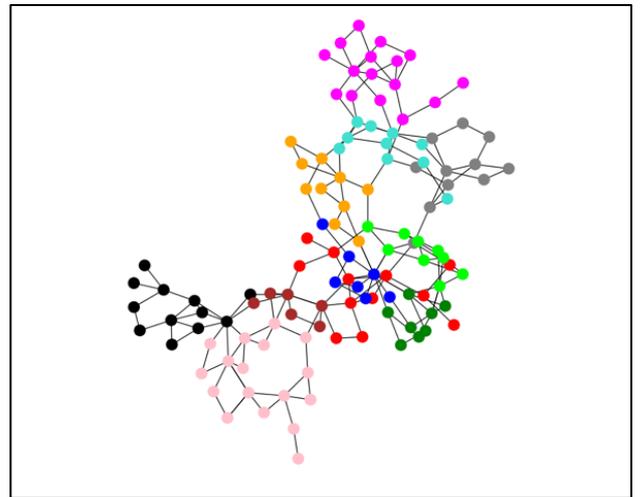

**Fig. 6.** Partitioned Graph with the highest modularity 0.819572, with group number k=11, by Fujitsu Digital Annealer

## IV. CONCLUSION

In this study, we explored number partitioning and graph partitioning problems with the digital annealer (DA) developed by Fujitsu Ltd. These problems were first converted to Quadratic Unconstrained Binary Optimization (QUBO) forms and then solved by DA. As a direct analogy to the Ising model, number partitioning problems were explored on DA. It was shown that the optimal solutions were achieved in less than 30 seconds for 6500 numbers without the need of tuning hyperparameters. It evidences the convenience of DA for users. In our future work, we will adopt complex test cases and investigate the well-defined timing metric for heuristic optimization problems [28] on DA.

For Karate club, DA achieved a modularity of 0.445, higher than the optimal values obtained by quantum annealing and simulated annealing reported in the literature, to the best of our knowledge. For the partition of power distribution networks, we solved the test cases IEEE 33–bus and IEEE 118–bus graphs. DA identified communities efficiently. The partitioning result indicated that the nodes were connected more densely within the sub–communities, as expected by the definition of modularity. Our results demonstrate a lucid exemplification of the application of DA to optimizing power distribution systems among the industry.

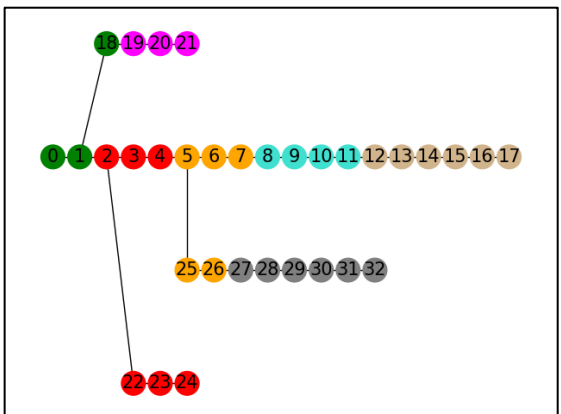

**Fig. 4.** Partitioned Graph with the highest modularity 0.743158, with group number k=7, by Fujitsu Digital Annealer

## ACKNOWLEDGEMENTS

H.C.H would like to acknowledge Prof. Yu–Cheng Lin and Yu–Chen Shu for their insightful discussions. This work is supported by NSTC under the grant No. 112–2119–M–006–004.